\documentclass{article}

\usepackage[utf8]{inputenc} \usepackage[T1]{fontenc}    \usepackage{hyperref}       \usepackage{url}            \usepackage{booktabs}       \usepackage{amsfonts}       \usepackage{nicefrac}       \usepackage{microtype}      \usepackage{xcolor}         \usepackage[a4paper, top=2.5cm, bottom=2.5cm, left=3cm, right=3cm]{geometry}

\usepackage{amsmath,amsthm,amssymb}
\usepackage[capitalize,noabbrev]{cleveref}
\usepackage[square,numbers]{natbib}
\usepackage{nicefrac}
\usepackage{tikz}
\usepackage{enumitem}
\usepackage{authblk}

\newtheorem{innerdefinition}{Definition}
\newtheorem*{innerdefinitionnonumber}{\definitionname}
\providecommand{\definitionname}{}

\usepackage{tikz}
\usetikzlibrary{arrows.meta,positioning,calc}
\usepackage{colortbl}

\newcommand{\lmultiset}{\mathopen{\{\mkern-4mu\vert}}
\newcommand{\rmultiset}{\mathclose{\vert\mkern-4mu\}}}

\usepackage{xparse}
\makeatletter
\NewDocumentEnvironment{definition}{o}
 {  \IfNoValueTF{#1}
    {\innerdefinition}    {     \renewcommand{\definitionname}{#1}     \def\@currentlabel{#1}     \innerdefinitionnonumber    } }
 {\IfNoValueTF{#1}{\endinnerdefinition}{\endinnerdefinitionnonumber}}
\makeatother

\theoremstyle{definition}
\newtheorem{example}{Example}

\DeclareMathOperator{\flow}{flow}
\DeclareMathOperator{\dist}{dist}

\usepackage{graphicx}
\usepackage{tikz}

\newlength{\HMgap}
\newlength{\HMblockgap}
\newcommand{\HeatmapLabelFont}{\footnotesize\bfseries}
\newcommand{\HeatmapGeneratorFont}{\small\bfseries}
\newcommand{\HeatmapColumnGap}{\columnsep}
\newcommand{\HeatmapPanelWidth}{16.4mm}
\newcommand{\HeatmapMaxHeight}{0.49\textheight}

\newlength{\HMtotalwidth}
\newlength{\HMnaturalwidth}
\newlength{\HMcolgap}
\newlength{\HMlabelwidth}
\newlength{\HMlabelheight}
\newlength{\HMallnodeswidth}
\newlength{\HMmethodheight}
\newlength{\HMcell}
\newlength{\HMstep}
\newlength{\HMgridwidth}
\newlength{\HMleftx}
\newlength{\HMrightx}
\newlength{\HMdividerx}
\newlength{\HMprobedheight}
\newlength{\HMtoprowheight}
\newlength{\HMtopheight}
\newlength{\HMrectrowheight}
\newlength{\HMrectheight}
\newlength{\HMrefheight}
\newlength{\HMrefcenter}
\newlength{\HMbottomtop}
\newlength{\HMbottomheight}
\newlength{\HMbottomend}
\newlength{\HMrecttop}
\newlength{\HMrowstep}
\newlength{\HMrowcenter}
\newlength{\HMpanelwidthlimit}
\newlength{\HMheightlimit}
\newsavebox{\HMprobe}
\newsavebox{\HMcombinedbox}

\newcommand{\HeatmapImage}[2]{%
  \includegraphics[width=#1]{#2}%
}

\newcommand{\HMMeasureImage}[2]{%
  \sbox{\HMprobe}{\HeatmapImage{\HMcell}{#1}}%
  \setlength{#2}{\dimexpr\ht\HMprobe+\dp\HMprobe\relax}%
}

\newcommand{\HMDrawResultGrid}[4]{%
  \begingroup%
  \setlength{\HMrowstep}{\dimexpr#4+\HMgap\relax}%
  \begin{scope}[xshift=#2]
    \foreach \centrality [count=\row from 0] in {
      degree,closeness,betweenness,pagerank%
    } {
      \setlength{\HMrowcenter}{%
        \dimexpr#3+\row\HMrowstep+#4/2\relax}%
      \foreach \aggregation [count=\col from 0] in {sum,min,max,pav} {
        \node[hm image] (hm-#1-\centrality-\aggregation)
          at (\col,-\HMrowcenter)
          {\HeatmapImage{\HMcell}{#1/\centrality_\aggregation.png}};
      }
    }
  \end{scope}
  \endgroup%
}

\newcommand{\HMDrawHeaders}[2]{%
  \coordinate (hm-#1-top) at (0pt,-#2);
  \foreach \aggregation/\heading in {sum/Sum,min/Min,max/Max,pav/Prop} {
    \node[hm label, anchor=south]
      at ([yshift=\HMgap]hm-#1-degree-\aggregation.center |- hm-#1-top)
      {\strut\heading};
  }
}

\newcommand{\HMDrawMethod}[3]{%
  \node[hm method, anchor=north]
    at ([yshift=-\HMgap]\dimexpr#2+\HMgridwidth/2\relax,-#3)
    {\strut #1};
}

\newcommand{\AllGroupHeatmaps}[1][\linewidth]{%
  \begingroup%
  \setlength{\HMtotalwidth}{#1}%
  \setlength{\HMcolgap}{\HeatmapColumnGap}%
  \sbox{\HMprobe}{{\HeatmapLabelFont\strut Betweenness}}%
  \setlength{\HMlabelwidth}{\wd\HMprobe}%
  \setlength{\HMlabelheight}{\dimexpr\ht\HMprobe+\dp\HMprobe\relax}%
  \sbox{\HMprobe}{{\HeatmapLabelFont\strut All Nodes}}%
  \setlength{\HMallnodeswidth}{\wd\HMprobe}%
  \sbox{\HMprobe}{{\HeatmapGeneratorFont\strut Two circles}}%
  \setlength{\HMmethodheight}{\dimexpr\ht\HMprobe+\dp\HMprobe\relax}%
  \setlength{\HMleftx}{\HMallnodeswidth}%
  \ifdim\HMleftx<\HMlabelheight
    \setlength{\HMleftx}{\HMlabelheight}%
  \fi%
  \addtolength{\HMleftx}{\HMgap}%
  \setlength{\HMcell}{%
    \dimexpr(\HMtotalwidth-\HMleftx-\HMcolgap-6\HMgap)/8\relax}%
  \setlength{\HMpanelwidthlimit}{\HeatmapPanelWidth}%
  \ifdim\HMcell>\HMpanelwidthlimit
    \setlength{\HMcell}{\HMpanelwidthlimit}%
  \fi%
  \setlength{\HMstep}{\dimexpr\HMcell+\HMgap\relax}%
  \setlength{\HMgridwidth}{\dimexpr4\HMcell+3\HMgap\relax}%
  \setlength{\HMrightx}{\dimexpr\HMleftx+\HMgridwidth+\HMcolgap\relax}%
  \setlength{\HMnaturalwidth}{\dimexpr\HMrightx+\HMgridwidth\relax}%
  \setlength{\HMdividerx}{\dimexpr\HMleftx+\HMgridwidth+\HMcolgap/2\relax}%
  \HMMeasureImage{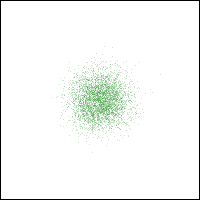}{\HMtoprowheight}%
  \HMMeasureImage{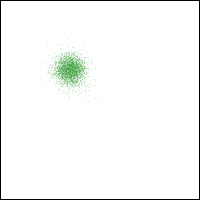}{\HMprobedheight}%
  \ifdim\HMtoprowheight<\HMprobedheight
    \setlength{\HMtoprowheight}{\HMprobedheight}%
  \fi%
  \ifdim\HMtoprowheight<\HMlabelwidth
    \setlength{\HMtoprowheight}{\HMlabelwidth}%
  \fi%
  \setlength{\HMtopheight}{\dimexpr4\HMtoprowheight+3\HMgap\relax}%
  \HMMeasureImage{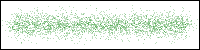}{\HMrectrowheight}%
  \ifdim\HMrectrowheight<\HMlabelheight
    \setlength{\HMrectrowheight}{\HMlabelheight}%
  \fi%
  \setlength{\HMrectheight}{\dimexpr4\HMrectrowheight+3\HMgap\relax}%
  \HMMeasureImage{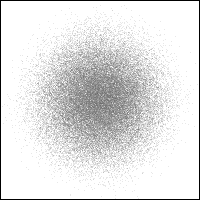}{\HMrefheight}%
  \HMMeasureImage{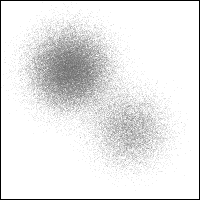}{\HMprobedheight}%
  \ifdim\HMrefheight<\HMprobedheight
    \setlength{\HMrefheight}{\HMprobedheight}%
  \fi%
  \HMMeasureImage{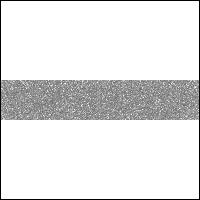}{\HMprobedheight}%
  \ifdim\HMrefheight<\HMprobedheight
    \setlength{\HMrefheight}{\HMprobedheight}%
  \fi%
  \ifdim\HMrefheight<\HMlabelheight
    \setlength{\HMrefheight}{\HMlabelheight}%
  \fi%
  \setlength{\HMbottomheight}{\HMrectheight}%
  \ifdim\HMbottomheight<\HMrefheight
    \setlength{\HMbottomheight}{\HMrefheight}%
  \fi%
  \setlength{\HMbottomtop}{\dimexpr\HMtopheight+\HMgap+\HMmethodheight
    +\HMblockgap+\HMlabelheight+\HMgap\relax}%
  \setlength{\HMrecttop}{%
    \dimexpr\HMbottomtop+(\HMbottomheight-\HMrectheight)/2\relax}%
  \setlength{\HMrefcenter}{\dimexpr\HMbottomtop+\HMbottomheight/2\relax}%
  \setlength{\HMbottomend}{\dimexpr\HMbottomtop+\HMbottomheight\relax}%
  \sbox{\HMcombinedbox}{%
    \begin{tikzpicture}[
      x=\HMstep,
      every node/.style={inner sep=0pt, outer sep=0pt},
      hm image/.style={anchor=west, draw=none},
      hm label/.style={font=\HeatmapLabelFont},
      hm method/.style={font=\HeatmapGeneratorFont}
    ]
      \path (0pt,0pt) rectangle (\HMnaturalwidth,0pt);
      \HMDrawResultGrid{Circle}{\HMleftx}{0pt}{\HMtoprowheight}%
      \HMDrawResultGrid{Circles}{\HMrightx}{0pt}{\HMtoprowheight}%
      \HMDrawHeaders{Circle}{0pt}%
      \HMDrawHeaders{Circles}{0pt}%
      \foreach \centrality/\heading in {
        degree/Degree,closeness/Closeness,
        betweenness/Betweenness,pagerank/PageRank%
      } {
        \node[hm label, rotate=90, anchor=south]
          at ([xshift=-\HMgap]hm-Circle-\centrality-sum.west)
          {\strut\heading};
        \node[hm label, rotate=90, anchor=south]
          at ([xshift=-\HMgap]hm-Circles-\centrality-sum.west)
          {\strut\heading};
      }
      \HMDrawMethod{Circle}{\HMleftx}{\HMtopheight}%
      \HMDrawMethod{Two circles}{\HMrightx}{\HMtopheight}%
      \HMDrawResultGrid{Rectangle}{\HMrightx}{\HMrecttop}{\HMrectrowheight}%
      \HMDrawHeaders{Rectangle}{\HMrecttop}%
      \foreach \centrality/\heading in {
        degree/Degree,closeness/Closeness,
        betweenness/Betweenness,pagerank/PageRank%
      } {
        \node[hm label, anchor=east]
          at ([xshift=-\HMgap]hm-Rectangle-\centrality-sum.west)
          {\strut\heading};
      }
      \HMDrawMethod{Rectangle}{\HMrightx}{\HMbottomend}%
      \coordinate (hm-reference-bottom) at (0pt,-\HMbottomend);
      \begin{scope}[xshift=\HMleftx]
        \foreach \folder/\heading [count=\col from 0] in {
          Circle/Circle,Circles/{Two circles},Rectangle/Rectangle%
        } {
          \node[hm image] (hm-ref-\folder) at (\col,-\HMrefcenter)
            {\HeatmapImage{\HMcell}{\folder/points.png}};
          \node[hm method, anchor=north]
            at ([yshift=-\HMgap]hm-ref-\folder.center |- hm-reference-bottom)
            {\strut\heading};
        }
        \node[hm label, rotate=90, anchor=south]
          at ([xshift=-\HMgap]hm-ref-Circle.west)
          {\strut All Nodes};
      \end{scope}
    \end{tikzpicture}%
  }%
  \setlength{\HMheightlimit}{\HeatmapMaxHeight}%
  \ifdim\dimexpr\ht\HMcombinedbox+\dp\HMcombinedbox\relax>\HMheightlimit
    \resizebox*{!}{\HMheightlimit}{\usebox{\HMcombinedbox}}%
  \else
    \usebox{\HMcombinedbox}%
  \fi%
  \endgroup%
}

\makeatletter
\@ifclassloaded{acmart}{%
  \newcommand{\PaperHeatmaps}{\AllGroupHeatmaps}%
}{
  \RequirePackage[T1]{fontenc}

  \newcommand{\PaperHeatmaps}{%
    \begingroup
      \linespread{1}\selectfont%

      \renewcommand{\HeatmapLabelFont}{%
        \fontencoding{T1}%
        \fontfamily{LinuxLibertineT-TLF}%
        \fontseries{b}\fontshape{n}%
        \fontsize{7}{8}\selectfont%
      }

      \renewcommand{\HeatmapGeneratorFont}{%
        \fontencoding{T1}%
        \fontfamily{LinuxLibertineT-TLF}%
        \fontseries{b}\fontshape{n}%
        \fontsize{8}{10}\selectfont%
      }%

      \renewcommand{\HeatmapColumnGap}{24pt}%
      \renewcommand{\HeatmapPanelWidth}{16.4mm}%
      \renewcommand{\HeatmapMaxHeight}{306.75pt}%
      
      \makebox[\linewidth][c]{%
        \hspace*{-4mm}%
        \AllGroupHeatmaps[150mm]%
        \hspace*{4mm}%
      }
    \endgroup
  }
}
\makeatother

\title{A Unified Framework for Extending\break Centrality Measures to Groups}

\author{Piotr Kępczyński}
\author{Oskar Skibski}
\affil{Institute of Informatics, University of Warsaw}

\begin{document}

\maketitle

\begin{abstract}
We propose a unified framework for extending centrality measures from individual nodes to groups. For each centrality measure, we introduce four group variants---maximum, minimum, sum, and proportional---which differ in how individual contributions are aggregated at the group level. The framework builds on a new unified perspective on major classes of centrality measures, according to which the score of a node can be expressed as a sum of assessments made by individual nodes, pairs of nodes, or groups of nodes. Each of the four variants then corresponds to a different rule for aggregating the assessments of the members of a group.
\end{abstract}

\section{Introduction}

Centrality measures are widely used across a range of research fields, including computer science, sociology, biology, economics, and engineering~\cite{Sciarra:etal:2018}.
A common approach to selecting a group of nodes for a given objective is to rank the nodes according to an individual centrality measure and choose the top-ranked ones.
Such strategies are used in network immunization~\cite{Chen:etal:2008,Altarelli:etal:2014}, in the identification of influential spreaders~\cite{Kitsak:etal:2010}, and in studies of the vulnerability of the Internet to targeted attacks~\cite{Albert:etal:2000,Cohen:etal:2001}.
However, selecting nodes solely according to their individual centrality scores ignores interactions and potential redundancy among the selected nodes, and may therefore lead to suboptimal groups~\cite{Morone:Makse:2015}.
For example, when promoting a product, selecting two influencers from different communities may provide a broader reach than selecting two individually more central influencers from the same one.

Group centrality measures address this limitation by evaluating groups of nodes as a whole.
Group variants have been defined for several well-known centrality measures, including degree~\cite{Gerbner:2022,Laussmann:2026}, closeness~\cite{Everett:Borgatti:1999}, betweenness~\cite{Brandes:2008,Everett:Borgatti:1999}, and PageRank~\cite{Liu:etal:2014}, as well as for vitality indices~\cite{Kucharczuk:Skibski:2026:groupvitality}.
However, existing approaches often rely on different underlying concepts, resulting in a fragmented landscape of group centrality definitions.
This makes it difficult to compare different group measures and systematically extend other node centralities to groups.

To address this, we propose a unified framework for extending node centrality measures to groups.
For each centrality measure expressed in our framework, we define four types of group extensions: \textsc{Sum}, \textsc{Max}, \textsc{Min}, and \textsc{Proportional}.
These extensions differ in how they aggregate the evaluations of individual group members provided by a particular node, pair of nodes, or group of nodes.
Our framework captures a wide range of existing group centrality measures as special cases, while also providing a systematic way to construct new variants (see Section~\ref{section:group_values}).

The following example illustrates how different group extensions of the same centrality measure can capture fundamentally different notions of group importance.

\begin{figure}[t]
\centering
\begin{tikzpicture}[x=7cm,y=14cm]
  \tikzset{
    house/.pic={
      \draw[thin]
        (-0.06,-0.04) -- (-0.06,0.01)
        -- (0.06,0.01) -- (0.06,-0.04) -- cycle;
      \draw[thin]
        (-0.08,0.000) -- (0,0.065) -- (0.08,0.00);

      \draw[thin]
        (-0.035,-0.04) -- (-0.035,0)
        -- (-0.005,0) -- (-0.005,-0.04);

      \draw[thin]
        (0.015,0.0) -- (0.015,-0.015)
        -- (0.040,-0.015) -- (0.040,-0.0) -- cycle;
      \draw[thin]
        (0.0275,0.0) -- (0.0275,-0.015);
      \draw[thin]
        (0.015,-0.00725) -- (0.040,-0.00725);

    }
  }

  \def\housesep{0.23}

  \foreach \i in {1,...,9} {
    \coordinate (\i) at ({(\i-1)*\housesep},0.65);
  }

  \draw[thick] (1) -- (9);

  \foreach \i in {1,...,9} {
    \fill[black] (\i) circle[radius=1.3pt];

    \pic at ($(\i)+(0,0.06)$) {house};

    \node[font=\scriptsize,inner sep=0pt]
      at ($(\i)+(0,0.09)$) {\i};
  }

\end{tikzpicture}
\caption{Nine buildings arranged along a street. Which three should be selected as locations for new stores?}
\label{figure:intro}
\end{figure}
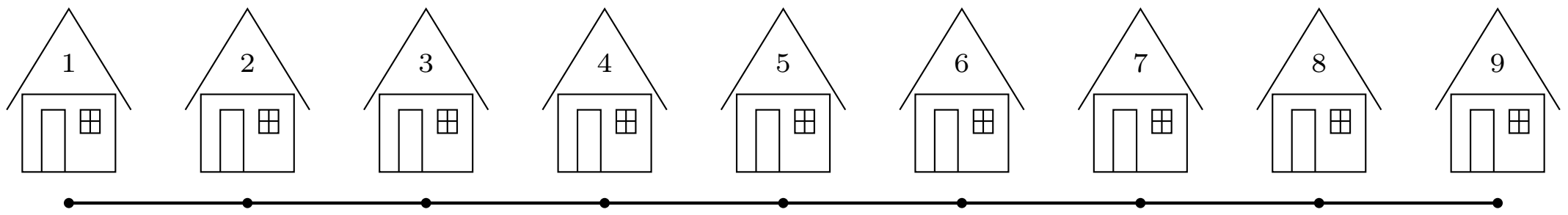

\begin{example}\label{example:intro}
Consider the problem of selecting locations for three stores along a street with nine buildings, as shown in Figure~\ref{figure:intro}.
A straightforward approach would be to select the three most central locations, $\{4,5,6\}$.
But is this always the best choice?

Following the intuition behind closeness centrality, assume that each node (building) evaluates a location by the negative of its distance to that location.
Let the overall score of a group of locations be the sum of the assessments it receives from all nodes.
How, then, should each node assess a group?

If the stores are of different types, each node may assess a group by summing its evaluations of all three locations.
Under \textsc{Sum} aggregation, the unique optimal group is $\{4,5,6\}$: the stores are placed in the three central buildings.
The same result is obtained if each node assesses a group based on its evaluation of the farthest store, as in \textsc{Min} aggregation.

However, if the stores are of the same type, each node may assess a group based on its evaluation of the closest store.
Under \textsc{Max} aggregation, the unique optimal group is $\{2,5,8\}$: each selected node has a store at distance zero, while every other node has a store at distance one.
Finally, nodes may care primarily about the closest store while also valuing alternatives in case that store is closed or a desired product is unavailable.
Under \textsc{Proportional} aggregation, each node evaluates the group as a weighted sum of its evaluations of the three locations, assigning weights of $1$, $1/2$, and $1/3$ to the closest, second closest, and farthest store, respectively.
The unique optimal group is then $\{3,5,7\}$, which yields an intermediate arrangement between the central locations preferred by \textsc{Sum} and \textsc{Min} and the more dispersed locations preferred by \textsc{Max}.

We provide detailed calculations in Example~\ref{example:group_intro}.
\hfill $\lrcorner$
\end{example}

Our framework builds on a new unified representation of node centrality measures, which we call \emph{SPEC}.
In this representation, a node's centrality is expressed as a sum of evaluations provided by individual nodes, pairs of nodes, or groups of nodes, possibly followed by a ranking-preserving conversion.
We show that a wide range of classical centrality measures admit such a representation, including distance-based, medial, feedback, and vitality measures.
While this representation is straightforward for some classes of centrality measures, others require more sophisticated interpretations.
In particular, for feedback centralities, we build on a walk-based interpretation~\cite{Was:Skibski:2021:feedback}, whereas for vitality indices, we use their game-theoretic characterization~\cite{Skibski:2021:vitality}.

The SPEC framework differs from existing approaches in its generality and purpose.
Previous work has classified centrality measures according to their structural properties~\cite{Borgatti:Everett:2006} or characterized them in terms of statistics based on paths, walks, and geodesics~\cite{Bloch:etal:2023}.
In contrast, SPEC allows for general evaluation functions and provides a common representation in terms of evaluations by sets of nodes, designed specifically to enable systematic extensions to groups.
Furthermore, SPEC follows a recent line of research connecting centrality analysis with social choice theory by viewing the centrality of a node or group as the aggregation of assessments provided by separate evaluators, analogous to voters~\cite{Skibski:2023:closeness,Laussmann:2026,Papasotiropoulos:etal:2026:proportional}.

We complement our theoretical framework with an experimental analysis of the different group extensions on illustrative graphs.
Our results demonstrate that the choice of aggregation rule can substantially affect which groups are considered optimal.
In particular, the \textsc{Proportional} extension can produce selections that better reflect the relative sizes of different communities.
For example, when selecting four nodes, it can favor allocating one of them to a community containing one quarter of the nodes in the graph.

\section{Preliminaries}

We focus on undirected, unweighted graphs. However, the framework and our results can also be extended to directed and weighted graphs.
A \emph{graph} $G=(V,E)$ consists of a set of nodes $V$ and a set of edges $E \subseteq \{\{u,v\} : u,v \in V,\ u \neq v\}$.
For an edge $\{u,v\}\in E$, nodes $u$ and $v$ are \emph{incident} to the edge and \emph{adjacent} to each other; they are also called \emph{neighbors}.
The number of incident edges of a node is its degree, denoted by $\deg(v)$.

A \emph{walk} is a sequence of nodes $p=(u_0,\dots,u_k)$, where $k\geq 0$, such that every pair of consecutive nodes is connected by an edge, i.e., $\{u_i,u_{i+1}\}\in E$ for every $i\in\{0,\dots,k-1\}$.
The length of a walk is the number of traversed edges, i.e., $k$.
The set of all walks in a graph, including walks of size zero, is denoted by $\Omega(G)$.
The \emph{distance} between nodes $u$ and $v$, denoted by $\dist_G(u,v)$, is the length of a shortest walk between them.
We set $\dist_G(u,v)=\infty$ if no such walk exists and omit the subscript $G$ when the graph is clear from the context.

By $M^X$ we denote the set of all nonempty finite multisets of $X$.
For $M\in M^X$, let $|M|$ denote its size, the number of elements in $M$.
We use the notation $\lmultiset \cdot \rmultiset$ to list elements of the multiset.

\section{SPEC representation}\label{section:spec_representation}

In this section, we introduce a unified representation, called \emph{SPEC}, which will allow us to extend node centrality measures to groups.

A \emph{SPEC representation} is a tuple $\mathcal{R}=(\mathcal{S}, P, e, c)$ consisting of the following components:
\begin{itemize}
    \item A \emph{selector} $\mathcal{S}$ that assigns to each finite set of nodes $W$ a family of subsets $\mathcal{S}(W)\subseteq 2^W$, called \emph{evaluators}.
    \item A \emph{policy} $P$ that determines whether nodes $S \subseteq V$ evaluate themselves. We consider two options:
    \begin{itemize}
        \item \emph{Exclusive policy} $P_{exc}(S) = S$: nodes from $S$ are excluded.
        \item \emph{Inclusive policy} $P_{inc}(S) = \emptyset$: no nodes are excluded.
    \end{itemize}
    \item An \emph{evaluation} $e_G:2^V\times V\rightarrow \mathbb{R}$.
    The value $e_G(U,v)$ specifies how an evaluator $U$ evaluates a node $v$.
    \item A \emph{conversion} $c:\mathbb{R}\rightarrow\mathbb{R}$ that is a strictly increasing function on the set of attainable aggregate evaluation scores and thus preserves the ranking of nodes.
\end{itemize}

The node centrality induced by a SPEC representation $\mathcal{R}=(\mathcal{S}, P, e, c)$ is defined for every graph $G=(V,E)$ and node $v\in V$ as
\[
F^{\mathcal{R}}_v(G)=c\left(\sum_{U\in \mathcal{S}(V \setminus P(\{v\}))}e_G(U,v)\right).
\]

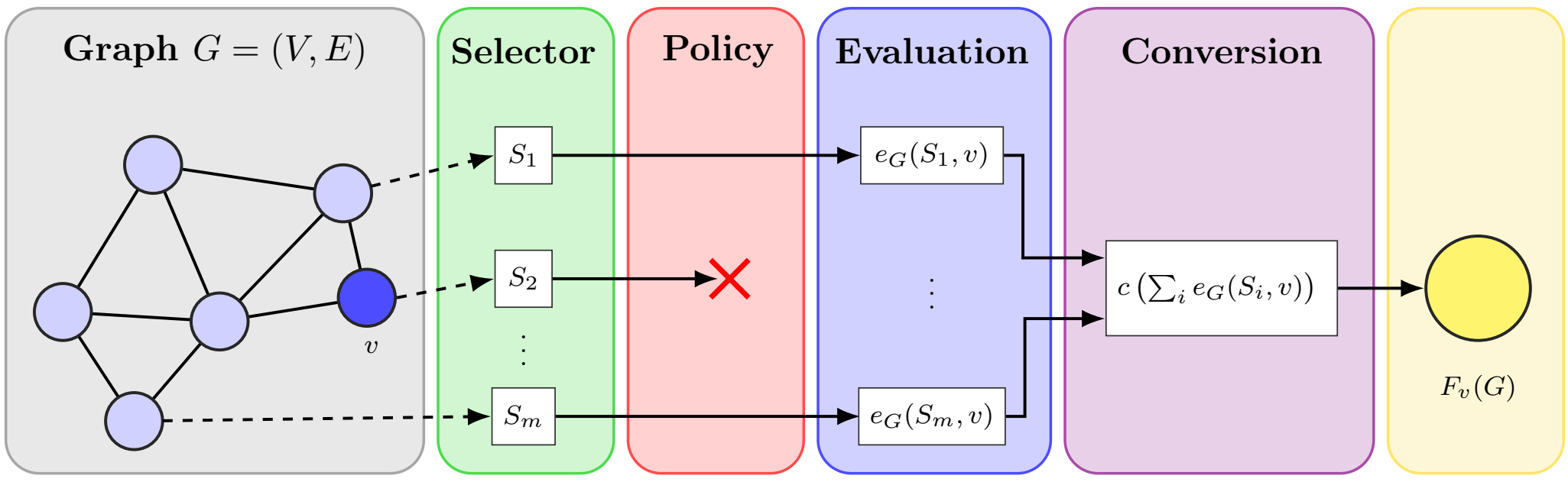
\begin{figure}[t]
\centering

\newcommand{\SpecFigureWidth}{0.9\linewidth}
\newcommand{\SpecTitleFont}{\fontsize{11}{12}\selectfont}
\newcommand{\SpecMathFont}{\fontsize{8}{10}\selectfont}

\resizebox{\SpecFigureWidth}{!}{%
\begin{tikzpicture}[
    >=Latex,
    font=\SpecMathFont,
    panel title/.style={
        font=\SpecTitleFont\bfseries
    },
    panel/.style={
        rounded corners=10pt,
        draw=#1!70,
        fill=#1!18,
        line width=0.8pt
    },
    smallbox/.style={
        draw=black!80,
        fill=white,
        minimum width=0.6cm,
        minimum height=0.6cm,
        align=center
    },
    formula box/.style={
        draw=black!80,
        fill=white,
        minimum width=2.3cm,
        minimum height=1cm,
        align=center
    },
    arr/.style={-Latex, line width=0.9pt},
    dashedarr/.style={-Latex, line width=0.9pt, dashed},
    evalarr/.style={-Latex, line width=0.9pt},
    crossed/.style={red, line width=1.6pt}
]

\coordinate (O) at (0,0);

\draw[panel=gray] (0,0) rectangle (4.4,4.9);
\node[panel title] at (2.2,4.45) {Graph $G=(V,E)$};

\draw[panel=green!80!black] (4.55,0) rectangle (6.4,4.9);
\node[panel title] at (5.45,4.45) {Selector};

\draw[panel=red] (6.55,0) rectangle (8.4,4.9);
\node[panel title] at (7.5,4.45) {Policy};

\draw[panel=blue] (8.55,0) rectangle (11.0,4.9);
\node[panel title] at (9.75,4.45) {Evaluation};

\draw[panel=violet] (11.15,0) rectangle (14.4,4.9);
\node[panel title] at (12.8,4.45) {Conversion};

\draw[panel=yellow!80!orange] (14.55,0) rectangle (16.4,4.9);

\tikzset{
    gnode/.style={
        circle, draw=black!85, fill=blue!18,
        minimum size=6mm, inner sep=0pt, line width=0.9pt
    },
    vnode/.style={
        circle, draw=black!85, fill=blue!70,
        minimum size=6mm, inner sep=0pt, line width=0.9pt
    },
    wnode/.style={
        circle, draw=black!85, fill=blue!32,
        minimum size=6mm, inner sep=0pt, line width=0.9pt
    }
}

\coordinate (g1) at (0.60,1.70);
\coordinate (g2) at (1.55,3.25);
\coordinate (g3) at (2.25,1.60);
\coordinate (g4) at (3.55,2.95);
\coordinate (gv) at (3.80,1.85);
\coordinate (g6) at (1.35,0.55);

\draw[line width=0.9pt]
    (g1) -- (g2)
    (g1) -- (g6)
    (g1) -- (g3)
    (g2) -- (g3)
    (g2) -- (g4)
    (g3) -- (g4)
    (g3) -- (g6)
    (g3) -- (gv)
    (g4) -- (gv);

\node[gnode] at (g1) {};
\node[gnode] at (g2) {};
\node[gnode] at (g3) {};
\node[gnode] at (g4) {};
\node[vnode] (vnode) at (gv) {};
\node[gnode] at (g6) {};

\node at ($(gv)+(0.05,-0.52)$) {$v$};

\node[smallbox] (S1) at (5.45,3.35) {$S_1$};
\node[smallbox] (S2) at (5.45,2.05) {$S_2$};
\node at (5.45,1.4) {$\vdots$};
\node[smallbox] (Sm) at (5.45,0.60) {$S_m$};

\draw[dashedarr, shorten <=3mm] (g4) -- (S1.west);
\draw[dashedarr] (vnode.east) -- (S2.west);
\draw[dashedarr, shorten <=3mm] (g6) -- (Sm.west);

\draw[arr] (S1.east) -- (9.0,3.35);
\draw[arr] (Sm.east) -- (9.0,0.60);

\draw[arr] (S2.east) -- (7.50,2.05);
\draw[crossed] (7.42,1.85) -- (7.82,2.25);
\draw[crossed] (7.42,2.25) -- (7.82,1.85);

\node[smallbox, minimum width=1.5cm]
    (E1) at (9.75,3.35) {$e_G(S_1,v)$};

\node at (9.75,2.00) {$\vdots$};

\node[smallbox, minimum width=1.5cm]
    (Em) at (9.75,0.60) {$e_G(S_m,v)$};

\node[formula box] (Conv) at (12.8,1.95) {%
    $\textstyle
    c\left(\sum_{i} e_G(S_i,v)\right)$
};

\draw[arr] (E1.east) -- ++(0.20,0)
    |- ([yshift=0.32cm]Conv.west);

\draw[arr] (Em.east) -- ++(0.20,0)
    |- ([yshift=-0.32cm]Conv.west);

\node[
    circle,
    draw=black!85,
    fill=yellow!70,
    minimum size=11mm,
    line width=0.9pt
] (score) at (15.50,1.95) {};

\draw[arr] (Conv.east) -- (score.west);
\node at (15.5,0.9) {$F_v(G)$};

\end{tikzpicture}%
}

\caption{Illustration of the SPEC framework. The selector chooses
candidate evaluators, the policy may exclude some of them, the
remaining evaluators assign scores to node $v$, and the resulting
aggregate score is converted into the final centrality value.}
\label{figure:spec}
\end{figure}

Under this representation, a node $v$ is evaluated by a family of node sets selected by the selector $\mathcal{S}$.
Depending on the policy $P$, these sets may or may not contain $v$ itself.
Each selected set, called an \emph{evaluator}, assigns a value to $v$ according to the evaluation function $e_G$.
The resulting values are summed and then transformed by the conversion function $c$.
See Figure~\ref{figure:spec} for an illustration.

Appendix~\ref{appendix:spec} provides SPEC representations of various classical centrality measures.
Below, we discuss a few illustrative examples.

\subsubsection*{Degree centrality}\label{section:degree}
Degree centrality simply counts the number of edges incident to a node:
\[
F_v(G) = |\{\{u,v\} \in E : u \in V\}|.
\]
To express it in the SPEC framework, we let each node other than $v$ assign a score of $1$ to $v$ if they are connected, and $0$ otherwise.
\begin{definition}\label{definition:spec_degree}
A SPEC representation of degree centrality is:
\begin{itemize}
    \item $\mathcal{S}(W) = \{\{w\} : w \in W\}$ (individual nodes are evaluators);
    \item $P = P_{exc}$ (a node does not evaluate itself);
    \item $e_G(\{w\}, v) = 1$ if $\{w,v\} \in E$, and $0$ otherwise (node $v$ receives $1$ from each neighbor and $0$ from every other node);
    \item $c(r) = r$ (no conversion).
\end{itemize}
\end{definition}

\subsubsection*{Closeness centrality}\label{section:closeness}
Closeness centrality~\cite{Bavelas:1948} is a centrality measure typically defined for connected graphs. It assesses a node based on the sum of its distances to all other nodes:
\[
F_v(G)=\dfrac{1}{\sum_{u\in V\setminus\{v\}}\dist(u,v)}.
\]
Unlike degree centrality, closeness centrality is not directly expressed as a sum of individual evaluations.
However, we can represent it as a sum of negative distances to other nodes, followed by a conversion function that transforms this sum into its reciprocal.
\begin{definition}\label{definition:spec_closeness}
A SPEC representation of closeness centrality is:
\begin{itemize}
    \item $\mathcal{S}(W) = \{\{w\} : w \in W\}$ (individual nodes are evaluators);
    \item $P = P_{exc}$ (a node does not evaluate itself);
    \item $e_G(\{w\}, v) = -\dist(w,v)$ (node $w$ evaluates node $v$ by the negative of the distance between them);
    \item $c(r) = -1/r$ (converts the negative sum of distances into the reciprocal of their sum).
\end{itemize}
\end{definition}

\subsubsection*{Betweenness centrality}\label{section:betweenness}
Betweenness centrality~\cite{Freeman:1977} measures how often a node lies on shortest paths between other nodes.
It is defined as
\[
F_v(G)=\sum_{\{s,t\}\subseteq V\setminus\{v\}, s\neq t}\dfrac{\sigma_{st}(v)}{\sigma_{st}},
\]
where $\sigma_{st}$ is the number of shortest paths between nodes $s$ and $t$, and $\sigma_{st}(v)$ is the number of such paths that pass through node $v$.
If $\sigma_{st}=0$, the corresponding fraction is defined to be $0$.

Betweenness centrality can be expressed in the SPEC framework by using pairs of nodes as evaluators.
\begin{definition}\label{definition:spec_betwennness}
A SPEC representation of betweenness centrality is:
\begin{itemize}
    \item $\mathcal{S}(W) = \{\{s,t\} \subseteq W : s \neq t\}$ (pairs of nodes are evaluators);
    \item $P = P_{exc}$ (a node does not evaluate itself);
    \item $e_G(\{s,t\}, v) = \sigma_{st}(v)/\sigma_{st}$ (a pair of nodes $s,t$ evaluates node $v$ by the proportion of shortest paths between them that pass through $v$);
    \item $c(r) = r$ (no conversion).
\end{itemize}
\end{definition}

\subsubsection*{PageRank}\label{section:pagerank}
Unlike centrality measures based on shortest paths, PageRank~\cite{Page:etal:1999} takes into account walks of all lengths and measures the importance of a node based on their weighted contributions.
PageRank has several variants; one standard version is defined as follows:
\[
F_v(G) = \frac{1-\alpha}{|V|}
\sum_{\substack{(u_0,\dots,u_k)\in\Omega(G)\\u_k=v}}
\frac{\alpha^k}{\prod_{i=0}^{k-1}\deg(u_i)},
\]
where $\alpha\in(0,1)$ is a fixed \emph{damping factor}.

A natural way to decompose this formula is to group walks by their starting nodes.
This approach is also closely related to personalized PageRank, discussed in the original paper~\cite{Page:etal:1999}, where only selected nodes may have a positive probability of being chosen as starting points.
In this way, each node evaluates other nodes based on the weighted contributions of walks starting at that node.
Importantly, walks starting at the evaluated node must also be included.

\begin{definition}\label{definition:spec_pagerank}
A SPEC representation of PageRank is:
\begin{itemize}
    \item $\mathcal{S}(W) = \{\{w\} : w \in W\}$ (individual nodes are evaluators);
    \item $P = P_{inc}$ (a node also evaluates itself);
    \item $e_G(\{w\},v) = \frac{1-\alpha}{|V|}
    \sum_{\substack{(u_0,\dots,u_k)\in\Omega(G)\\u_0=w,\ u_k=v}}
    \frac{\alpha^k}{\prod_{i=0}^{k-1}\deg(u_i)}$
    (node $w$ evaluates node $v$ by the total weighted contribution of walks from $w$ to $v$);
    \item $c(r) = r$ (no conversion).
\end{itemize}
\end{definition}

See \cref{figure:numerical_example} for a numerical example comparing degree, closeness and PageRank centralities.

\begin{figure}[t]
    \centering
        \begin{tikzpicture}

\tikzset{
    gnode/.style={circle,draw=black!85,fill=blue!18,minimum size=6mm,inner sep=0pt,line width=0.9pt},
    vnode/.style={circle,draw=black!85,fill=blue!70,minimum size=6mm,inner sep=0pt,line width=0.9pt},
    wnode/.style={circle,draw=black!85,fill=blue!32,minimum size=6mm,inner sep=0pt,line width=0.9pt}
}

\coordinate (g1) at (0.60,1.70);
\coordinate (g2) at (1.55,3.25);
\coordinate (g3) at (2.25,1.60);
\coordinate (g4) at (3.55,2.95);
\coordinate (gv) at (3.80,1.85);
\coordinate (g6) at (1.35,0.55);

\draw[line width=0.9pt]
    (g1) -- (g2)
    (g1) -- (g6)
    (g1) -- (g3)
    (g2) -- (g3)
    (g2) -- (g4)
    (g3) -- (g4)
    (g3) -- (g6)
    (g3) -- (gv)
    (g4) -- (gv);

\node[gnode] at (g1) {$3$};
\node[gnode] at (g2) {$1$};
\node[gnode] at (g3) {$4$};
\node[gnode] at (g4) {$2$};
\node[vnode,text=white] (vnode) at (gv) {$5$};
\node[gnode] at (g6) {$6$};

        \node[anchor=west,inner sep=0pt] at (5,2) {%
            \renewcommand{\arraystretch}{1.1}%
            \setlength{\tabcolsep}{8pt}%
            \begin{tabular}{@{}lccc@{}}
                \toprule
                Evaluators & Degree & Closeness &
                \begin{tabular}{@{}c@{}}
                    PageRank
                    $(\alpha=0.85)$
                \end{tabular}\\
                \midrule
                $\{1\}$ & $0$ & $-2$ & $0.0141$\\
                $\{2\}$ & $1$ & $-1$ & $0.0204$\\
                $\{3\}$ & $0$ & $-2$ & $0.0124$\\
                $\{4\}$ & $1$ & $-1$ & $0.0171$\\
                $\{5\}$ & \textemdash & \textemdash & $0.0410$\\
                $\{6\}$ & $0$ & $-2$ & $0.0125$\\
                \midrule
                Sum $r$
                    & $2$ & $-8$ & $0.1176$\\
                Conversion $c(r)$
                    & $2$ & $1/8$ & $0.1176$\\
                \bottomrule
            \end{tabular}%
        };
\end{tikzpicture}%
\caption{SPEC evaluations of node $5$ for degree, closeness, and
PageRank centrality. A dash indicates that the evaluator is excluded
by the exclusive policy. The conversion is $c(r)=-1/r$ for closeness
and $c(r)=r$ for degree and PageRank. PageRank values are rounded to
four decimal places.}
\label{figure:numerical_example}
\end{figure}

We note that a centrality measure may admit multiple SPEC representations.
Different representations may reflect different perspectives on and interpretations of the same centrality measure.
In the following sections, we show how the main classes of centrality measures can be represented within the SPEC framework.

\subsection{Distance-based centralities}\label{section:distance-based}
A centrality measure is \emph{distance-based}~\cite{Skibski:Sosnowska:2018:distance} (or \emph{geometric}~\cite{Boldi:etal:2025}) if the importance of a node depends only on its distances to other nodes.
Most distance-based centralities are additive and can be written in the form
\[
F_v(G)=\sum_{u\in V\setminus\{v\}}f(\dist(u,v))
\]
for some non-increasing function $f:\mathbb{N}_{>0}\cup\{\infty\}\rightarrow\mathbb{R}$, where $f(\infty)$ specifies the contribution of unreachable nodes.

This class includes many classical centralities, such as degree (with $f(1)=1$ and $f(x)=0$, otherwise), harmonic (with $f(x)=1/x$), decay (with $f(x)=\alpha^x$ for a fixed $\alpha\in(0,1)$), and $k$-step reach (with $f(x) = 1$ if $x \le k$ and $f(x)=0$, otherwise).
For these four measures, we set $f(\infty)=0$.
A prominent non-additive example is closeness centrality, which, on connected graphs with at least two nodes, induces the same ranking as the additive measure obtained by setting $f(x)=-x$.
This additive form was used in Example~\ref{example:intro} and underlies the SPEC representation in Definition~\ref{definition:spec_closeness}.

Additive distance-based centralities admit a natural SPEC representation.
Individual nodes serve as evaluators ($\mathcal{S}(W)=\{\{u\}:u\in W\}$), and the exclusive policy is used ($P=P_{exc}$).
Each node $u$ evaluates a node $v\neq u$ according to $f$ applied to their distance ($e_G(\{u\},v)=f(\dist(u,v))$), and no conversion is required ($c(r)=r$).

\subsection{Medial centralities}
Another major class of centrality measures consists of \emph{medial} (or \emph{betweenness-like}) centralities.
These measures assess nodes according to their role as intermediaries between other nodes.
To this end, they consider pairs of distinct nodes $s,t$ and quantify the contribution of the evaluated node $v$ to connecting them.
For variants that exclude pairs containing $v$, this can be expressed as
\[
F_v(G)=\sum_{\{s,t\} \subseteq V\setminus\{v\}, s \neq t}\Delta^{s,t}_v(G),
\]
where $\Delta^{s,t}_v(G)$ denotes the contribution of $v$ to the pair $\{s,t\}$.

This class includes classical betweenness centrality ($\Delta^{s,t}_v(G)=\sigma_{st}(v)/\sigma_{st}$) and stress centrality ($\Delta^{s,t}_v(G)=\sigma_{st}(v)$), as well as variants based on network flows, such as flow betweenness~\cite{Freeman:etal:1991}, and random walks~\cite{Bloechl:etal:2011,Newman:2005}.

Measures of this form admit a natural SPEC representation.
Pairs of distinct nodes serve as evaluators ($\mathcal{S}(W)=\{\{s,t\}:s,t\in W,\ s\neq t\}$), and the exclusive policy is used ($P=P_{exc}$).
A pair $\{s,t\}$ evaluates node $v$ according to $\Delta$ ($e_G(\{s,t\},v)=\Delta^{s,t}_v(G)$), and no conversion is required ($c(r)=r$).
Variants that also include pairs containing the evaluated node can instead be represented using the inclusive policy ($P=P_{inc}$).

Medial centralities can also be expressed in the SPEC representation, with paths rather than pairs of nodes serving as evaluators. 
We will discuss this formulation in more detail in Section~\ref{section:group_values}.

\subsection{Feedback centrality measures}
Feedback centralities~\cite{Was:Skibski:2021:feedback} constitute one of the classical classes of centrality measures.
These measures assess the importance of a node based on the number and importance of its neighbors.
Most of them have a recursive definition of the form
\[
F_v(G) = a(G) \sum_{\{u,v\} \in E} \frac{(F_u(G))^c}{(\deg(u))^d} + b(G),
\]
where $a$ and $b$ are real-valued functions defined on graphs, and $c,d\in\{0,1\}$.
This form captures several classical centrality measures:
\begin{itemize}
    \item degree centrality ($a(G)=1$, $b(G)=c=d=0$),
    \item beta measure~\cite{Brink:Gilles:1994} ($a(G)=d=1$, $b(G)=c=0$),
    \item eigenvector centrality~\cite{Bonacich:1972} ($a(G)=1/\lambda(G)$, $b(G)=d=0$, $c=1$),
    \item Seeley index~\cite{Seeley:1949} ($a(G)=c=d=1$, $b(G)=0$),
    \item Katz centrality~\cite{Katz:1953} ($c=1$, $d=0$, $a(G) \in (0,1/\lambda(G))$, $b(G)>0$), and
    \item PageRank~\cite{Page:etal:1999} ($c=d=1$, $a(G)=\alpha\in(0,1)$, $b(G)=(1-\alpha)/|V|$),
\end{itemize}
where $\lambda(G)$ denotes the largest eigenvalue of the adjacency matrix.

For degree centrality and the beta measure, individual neighbors naturally serve as evaluators, assigning a score of $1$ and $1/\deg(u)$, respectively, to a node $v$ adjacent to $u$ (see Definition~\ref{definition:spec_degree}).
For other feedback centralities, however, assuming that only neighbors evaluate a node based on their own importance is less natural.
First, the importance of each neighbor also depends on other nodes in the graph, whose contributions would not be represented directly.
Second, each evaluator would assign the same score to all its neighbors, which limits the distinctions captured by group aggregation.

To address this, we use the walk-based interpretation of feedback centralities presented by \citet{Was:Skibski:2021:feedback}.
Specifically, individual nodes serve as evaluators ($\mathcal{S}(W)=\{\{u\}:u\in W\}$), and the inclusive policy is used ($P=P_{inc}$).
Each node $u$ evaluates a node $v$ by the total weighted contribution of walks from $u$ to $v$ (for Katz centrality and PageRank) or by the limiting average of these contributions (for eigenvector centrality and the Seeley index).
The weights assigned to walks depend on the centrality measure.
No conversion is required ($c(r)=r$).
See Appendix~\ref{appendix:spec} for details.

\color{black}
\subsection{Vitality indices}

Vitality indices~\cite{Brandes:Erlebach:2005,Everett:Borgatti:2010} evaluate nodes according to the change in a graph's potential caused by their removal.
This class consists of measures of the form
\[
F_v(G)=p(G)-p(G-v)
\]
for some real-valued potential function $p$ defined on graphs, where $G-v$ denotes the graph obtained by removing $v$ and its incident edges.
This class includes degree centrality, flow betweenness centrality~\cite{Freeman:etal:1991} and attachment centrality~\cite{Skibski:etal:2019:attachment}.

It is not immediately clear how to express vitality indices in terms of evaluations of a node $v$ by groups of nodes.
However, such a representation follows from a game-theoretic perspective.
Specifically, \citet{Skibski:2021:vitality} proved that every vitality index admits the following Shapley-value representation:
\[
F_v(G)=\sum_{U\subseteq V} \beta(|V|,|U|) \left(\sum_{u\in U\cup\{v\}} F_u(G[U\!\cup\!\{v\}])-\sum_{u\in U} F_u(G[U])\right),
\]
where $\beta(n,k)=k!(n-k-1)!/n!$ for $k\in\{0,\dots,n-1\}$ and $\beta(n,n)=1$, and $G[S]$ is the subgraph of $G$ induced by $S$.
In this representation, each group $U\subseteq V\setminus\{v\}$ evaluates node $v$ by its weighted marginal contribution to the total centrality in the subgraph induced by $U\cup\{v\}$.
Thus, the contribution to the potential of the entire graph is expressed as a weighted sum of contributions to total centrality in induced subgraphs.

This yields the following SPEC representation of vitality indices.
All subsets of nodes serve as evaluators ($\mathcal{S}(W)=2^W$), and the inclusive policy is used ($P=P_{inc}$).
For $U\subseteq V\setminus\{v\}$, the evaluation is given by
\[
e_G(U,v)=\beta(|V|,|U|)\left(\sum_{u\in U\cup\{v\}} F_u(G[U\cup\{v\}])-\sum_{u\in U} F_u(G[U])\right).
\]
No conversion is required ($c(r)=r$).

\section{Group Extensions of SPEC}

In this section, we present four ways of extending a centrality measure given by a SPEC representation to a group centrality measure.
When a group rather than a single node is evaluated, each evaluator assigns a separate value to each node in the group.
The four variants differ in how these values are \emph{aggregated}.

Formally, an \emph{aggregation} is a function $A:M^\mathbb{R}\rightarrow\mathbb{R}$ such that $A(\lmultiset x\rmultiset)=x$ for every $x\in\mathbb{R}$.
We consider four aggregations:
\begin{itemize}
    \item \textsc{Sum} sums all evaluations: $A^{\mathrm{sum}}(M)=\sum_{x\in M}x$.
    \item \textsc{Max} selects the highest evaluation: $A^{\mathrm{max}}(M)=\max_{x\in M}x$.
    \item \textsc{Min} selects the lowest evaluation: $A^{\mathrm{min}}(M)=\min_{x\in M}x$.
    \item \textsc{Proportional} (\textsc{Prop} for short) takes a harmonically weighted sum of the evaluations, ordered from highest to lowest: $A^{\mathrm{prop}}(M)=\sum_{i=1}^{|M|}\nicefrac{x_i}{i}$, where $\vec{M}=(x_1,\dots,x_{|M|})$ lists the elements of $M$ in non-increasing order.
\end{itemize}

Given a SPEC representation $\mathcal{R}=(\mathcal{S},P,e,c)$ and an aggregation $A$, we define the corresponding group centrality for a nonempty group $S\subseteq V$ by
\[
F^{\mathcal{R},A}_{S}(G)=c\left(\sum_{U\in\mathcal{S}(V\setminus P(S))}A\left(\lmultiset e_G(U,v)\rmultiset_{v\in S}\right)\right).
\]
Each admissible evaluator $U$ evaluates every node in $S$, and $A$ combines these values into a single evaluation of the group.
These group evaluations are then summed over all admissible evaluators, and the conversion function $c$ is applied to the resulting sum.
Since $A(\lmultiset x\rmultiset)=x$, the group centrality agrees with the original node centrality on singletons: $F^{\mathcal{R},A}_{\{v\}}(G)=F^{\mathcal{R}}_v(G)$.

Before we proceed with the analysis of specific group measures, let us provide some intuition about how the different aggregation rules work.

\subsubsection*{\textsc{Sum} aggregation.}
The \textsc{Sum} aggregation simply sums all evaluations.
This makes the resulting group centrality closely related to the sum of the individual node centralities.
In fact, if the conversion is a positive linear scaling $c(x)=\alpha x$ for some $\alpha\in\mathbb{R}_{>0}$, which is often the case, and the inclusive policy $P_{inc}$ is used, the resulting group centrality is exactly equal to the sum of the individual node centralities:
\[
F^{\mathcal{R},A^{sum}}_{S}(G) =\sum_{v\in S} \alpha\sum_{U\in\mathcal{S}(V)}e_G(U,v) =\sum_{v\in S}F^{\mathcal{R}}_v(G).
\]

\subsubsection*{\textsc{Max} aggregation.}
The \textsc{Max} aggregation retains only the highest evaluation.
Thus, adding a new node to a group affects the aggregated evaluation of an evaluator only if the new node receives a higher evaluation than all nodes already in the group.
Consequently, the \textsc{Max} aggregation tends to reward spreading the nodes of the group across the graph, so that many evaluators have a well-evaluated representative in the group.

The \textsc{Max} aggregation is also closely related to the approach of merging the group into a single node and then evaluating the resulting node using the original centrality measure, which we call the \emph{merge approach}.
For example, for additive distance-based centralities whose evaluation function $f$ is non-increasing in the distance, the \textsc{Max} aggregation coincides with the merge approach.
Indeed, the distance from a node to the merged group is equal to its distance to the closest node in the group in the original graph.
The merge approach has been criticized for modifying the structure of the graph, which may lead to unintended effects for some centrality measures~\cite{Everett:Borgatti:1999}.
For example, in the graph
\tikz[baseline=-0.5ex, x=0.45cm,
    every node/.style={circle,draw,minimum size=3mm,inner sep=0pt,font=\scriptsize}]{
    \foreach \i in {1,2,3,6,7,8}
        \node (n\i) at ({\i-1+(\i>5 ? 0.2 : 0)},0) {\i};
    \node[fill=blue!15] (n4) at (3,0) {4};
    \node[fill=blue!15] (n5) at (4.2,0) {5};
    \draw (n1)--(n2)--(n3)--(n4);
    \draw (n5)--(n6)--(n7)--(n8);
}
nodes $4$ and $5$ both have zero betweenness centrality, yet merging them produces a node with betweenness $9$, artificially connecting two disconnected components.
The \textsc{Max} aggregation operates on the original graph and therefore avoids this issue.

\subsubsection*{\textsc{Min} aggregation.}
The \textsc{Min} aggregation retains only the lowest evaluation.
Thus, if even one node in the group receives a low evaluation from an evaluator, the aggregated evaluation assigned by that evaluator is also low.
Consequently, the \textsc{Min} aggregation tends to favor compact groups whose members receive relatively similar evaluations, rather than groups containing nodes that are far apart.

\subsubsection*{\textsc{Prop} aggregation.}
The \textsc{Prop} aggregation takes a harmonically weighted sum of the evaluations.
It often produces behavior intermediate between the \textsc{Sum} and \textsc{Max} aggregations: all evaluations are taken into account, but lower-ranked evaluations receive progressively smaller weights.
In fact, for groups of size two, the \textsc{Prop} aggregation is exactly the arithmetic mean of the \textsc{Max} and \textsc{Sum} aggregations.
The proportional nature of this aggregation results from diminishing returns when additional group members receive high evaluations from the same evaluators.
To illustrate this, suppose that nodes are partitioned into disconnected communities and that each evaluator assigns value $1$ to every node in its own community, including itself, and $0$ to every node outside it.
If a group contains $k$ nodes from a community of size $m$, then this community contributes $m (1+\frac{1}{2}+\cdots+\frac{1}{k})$ to the group score.
Adding another node from the same community therefore increases the score by $m/(k+1)$.
Consequently, an optimal group tends to allocate its members among communities proportionally to their sizes, reproducing the diminishing-returns principle underlying PAV in social choice theory~\cite{Lackner:Skowron:2023}.

\begin{example}\label{example:group_intro}
Recall Example~\ref{example:intro}.
Our goal is to extend the assessment inspired by closeness centrality to groups, using evaluations $e_G(\{w\},v)=-\dist(w,v)$ and the conversion $c(r)=-1/r$.
Table~\ref{table:group_intro} presents the scores obtained using all four aggregation rules under both policies.

\begin{table}[t]
\centering
\renewcommand{\arraystretch}{1.4}
\setlength{\tabcolsep}{4.5pt}
\begin{tabular}{@{}c|cccc|cccc@{}}
\toprule
& \multicolumn{4}{c}{Inclusive Closeness} & \multicolumn{4}{|c}{Exclusive Closeness} \\
\cmidrule(lr){1-1}\cmidrule(lr){2-5}\cmidrule(lr){6-9}
Group\  & \textsc{Sum} & \textsc{Min} & \textsc{Max} & \textsc{Prop} & \textsc{Sum} & \textsc{Min} & \textsc{Max} & \textsc{Prop} \\
\midrule
$\{4,5,6\}$ & \cellcolor{blue!10}$\nicefrac{1}{62}$ & \cellcolor{blue!10}$\nicefrac{1}{29}$ & $\nicefrac{1}{12}$ & $\nicefrac{6}{193}$ & $\nicefrac{1}{54}$ & \cellcolor{blue!10}$\nicefrac{1}{24}$ & $\nicefrac{1}{12}$ & $\nicefrac{1}{29}$ \\
$\{3,5,7\}$ & $\nicefrac{1}{68}$ & $\nicefrac{1}{38}$ & $\nicefrac{1}{8}$ & \cellcolor{blue!10}$\nicefrac{3}{95}$ & \cellcolor{blue!10}$\nicefrac{1}{52}$ & $\nicefrac{1}{28}$ & $\nicefrac{1}{8}$ & $\nicefrac{3}{76}$ \\
$\{2,5,8\}$ & $\nicefrac{1}{78}$ & $\nicefrac{1}{47}$ & \cellcolor{blue!10}$\nicefrac{1}{6}$ & $\nicefrac{6}{205}$ & $\nicefrac{1}{54}$ & $\nicefrac{1}{32}$ & \cellcolor{blue!10}$\nicefrac{1}{6}$ & \cellcolor{blue!10}$\nicefrac{3}{74}$ \\
$\{1,5,9\}$ & $\nicefrac{1}{92}$ & $\nicefrac{1}{56}$ & $\nicefrac{1}{8}$ & $\nicefrac{3}{122}$ & $\nicefrac{1}{60}$ & $\nicefrac{1}{36}$ & $\nicefrac{1}{8}$ & $\nicefrac{1}{28}$ \\
\bottomrule
\end{tabular}
\caption{Scores of the extensions to groups of the closeness centrality for selected groups from Figure~\ref{figure:intro}. 
Both versions of the policy are considered.
Shaded cells indicate the maximum over all three-node groups for the corresponding aggregation rule and policy.}
\label{table:group_intro}
\end{table}

In Example~\ref{example:intro}, we used the inclusive policy: a node (building) in which a new store is located also evaluates the group.
This is reasonable in our setting: the residents' access to stores still matters even when a store is located in their own building.
The \textsc{Sum} aggregation uniquely selects the group $\{4,5,6\}$, which minimizes the total distance from all nodes to the selected locations and thus maximizes the sum of evaluations.
The evaluations of this group by nodes $1,\dots,9$ are $(-12,-9,-6,-3,-2,-3,-6,-9,-12)$, which sum to $-62$ and yield a final score of $\nicefrac{1}{62}$ after conversion.
The \textsc{Min} aggregation also uniquely selects $\{4,5,6\}$: each node evaluates the group by the negative of its distance to the farthest selected location.
The resulting evaluations are $(-5,-4,-3,-2,-1,-2,-3,-4,-5)$, which sum to $-29$ and yield a final score of $\nicefrac{1}{29}$.
In contrast, the \textsc{Max} aggregation uniquely selects $\{2,5,8\}$, as each node evaluates the group by the negative of its distance to the closest selected location.
Every node therefore evaluates this group as either $0$ or $-1$: the evaluations are $(-1,0,-1,-1,0,-1,-1,0,-1)$, yielding a final score of $\nicefrac{1}{6}$.
Finally, the \textsc{Prop} aggregation uniquely selects $\{3,5,7\}$, which scores higher under this aggregation than either $\{4,5,6\}$ or $\{2,5,8\}$.
This aggregation accounts not only for the closest store but also for alternatives, with weights $(1,\nicefrac{1}{2},\nicefrac{1}{3})$ assigned in order of increasing distance.
The evaluations of $\{3,5,7\}$ by consecutive nodes are $(-6,-\nicefrac{25}{6},-\nicefrac{7}{3},-\nicefrac{5}{2},-\nicefrac{5}{3},-\nicefrac{5}{2},-\nicefrac{7}{3},-\nicefrac{25}{6},-6)$, which sum to $-\nicefrac{95}{3}$ and yield a final score of $\nicefrac{3}{95}$.

Alternatively, we can use the exclusive policy, under which the selected nodes no longer evaluate the group.
This may be appropriate if, for example, each selected building is entirely converted into a store and no residents remain there.
Under this policy, \textsc{Sum} uniquely selects $\{3,5,7\}$, with a final score of $\nicefrac{1}{52}$, while \textsc{Prop} uniquely selects $\{2,5,8\}$, with a final score of $\nicefrac{3}{74}$.
Both groups are more spread out than their respective optima under the inclusive policy.
This change arises because we omit the evaluations contributed by the selected nodes, whose values depend on the distances between the selected locations.
The \textsc{Min} aggregation still uniquely selects $\{4,5,6\}$, but its final score increases to $\nicefrac{1}{24}$.
The scores under \textsc{Max} remain unchanged for every group because each selected node has distance zero to the closest selected location and contributes $0$.
\hfill $\lrcorner$
\end{example}

\section{Group Values}\label{section:group_values}
In this section, we explore the group centrality measures resulting from our framework and their relationships to existing measures.

\subsubsection*{Group degree centrality}
According to Definition~\ref{definition:spec_degree}, each node $w$ outside the group $S$ assigns a score of $1$ to its neighbors in $S$ and $0$ to all other members of $S$.

Under the \textsc{Sum} aggregation, all these evaluations are summed, so the resulting group value equals the number of edges between $S$ and $V\setminus S$.
This corresponds to the size of the edge boundary of $S$, used for example by \citet{Li:Sheng:2020}.

Under the \textsc{Max} aggregation, a node $w$ evaluates the group $S$ by $1$ if at least one of its neighbors belongs to $S$, and by $0$ otherwise.
Hence, the group value equals the number of nodes outside the group that have at least one neighbor in the group.
This corresponds to the unnormalized group degree centrality~\cite{Everett:Borgatti:1999}.

Under the \textsc{Min} aggregation, a node $w$ evaluates the group by $1$ only if all members of $S$ are its neighbors.
Hence, the group value equals the number of common neighbors of all nodes in $S$.
This corresponds to the co-degree of $S$, considered by \citet{Gerbner:2022}.

Under the \textsc{Prop} aggregation, a node $w$ evaluates the group $S$ by the harmonic number $H_k=\sum_{i=1}^{k}1/i$, where $k$ is the number of its neighbors in $S$ and $H_0=0$.
This corresponds to the \emph{PAV group centrality index} of \citet{Laussmann:2026}.
See Figure~\ref{figure:group_numerical_example} for an example.

\begin{figure}[t]
    \centering
        \begin{tikzpicture}

\tikzset{
    gnode/.style={circle,draw=black!85,fill=blue!18,minimum size=6mm,inner sep=0pt,line width=0.9pt},
    vnode/.style={circle,draw=black!85,fill=blue!70,minimum size=6mm,inner sep=0pt,line width=0.9pt},
    wnode/.style={circle,draw=black!85,fill=blue!32,minimum size=6mm,inner sep=0pt,line width=0.9pt}
}

\coordinate (g1) at (0.50,1.70);
\coordinate (g2) at (1.25,3.25);
\coordinate (g3) at (1.9,1.60);
\coordinate (g4) at (2.85,2.95);
\coordinate (gv) at (3.15,1.85);
\coordinate (g6) at (1.1,0.55);

\draw[line width=0.9pt]
    (g1) -- (g2)
    (g1) -- (g6)
    (g1) -- (g3)
    (g2) -- (g3)
    (g2) -- (g4)
    (g3) -- (g4)
    (g3) -- (g6)
    (g3) -- (gv)
    (g4) -- (gv);

\node[vnode,text=white] (vnode) at (g1) {$3$};
\node[vnode,text=white] (vnode) at (g2) {$1$};
\node[gnode] at (g3) {$4$};
\node[gnode] at (g4) {$2$};
\node[gnode] at (gv) {$5$};
\node[gnode] at (g6) {$6$};

        \node[anchor=west,inner sep=0pt] at (4,2) {%
            \renewcommand{\arraystretch}{1.1}%
            \setlength{\tabcolsep}{6pt}%
            \begin{tabular}{@{}c|cc|cccc@{}}
                \toprule
                Evaluator $U$ & $e_G(U,1)$ & $e_G(U,3)$ & $A^{sum}$ & $A^{max}$ & $A^{min}$ & $A^{prop}$ \\
                \midrule
                $\{1\}$ & \textemdash & \textemdash & \textemdash & \textemdash & \textemdash & \textemdash\\
                $\{2\}$ & $1$         & $0$         & $1$ & $1$ & $0$ & $1$\\
                $\{3\}$ & \textemdash & \textemdash & \textemdash & \textemdash & \textemdash & \textemdash\\
                $\{4\}$ & $1$         & $1$         & $2$ & $1$ & $1$ & $\nicefrac{3}{2}$\\
                $\{5\}$ & $0$         & $0$         & $0$ & $0$ & $0$ & $0$\\
                $\{6\}$ & $0$         & $1$         & $1$ & $1$ & $0$ & $1$\\
                \midrule
                $F^{\mathcal{R}, A}_S(G)$
                    & & & $4$ & $3$ & $1$ & $\nicefrac{7}{2}$\\
                \bottomrule
            \end{tabular}%
        };
\end{tikzpicture}%
\caption{Group degree centrality evaluations of the group $S=\{1,3\}$ under four aggregation rules.
Under the exclusive policy, only nodes outside the group serve as evaluators.
Each evaluator assesses both nodes in the group, and these evaluations are aggregated.
Since no conversion is applied, the group centralities are the sums of the aggregated evaluations.}
\label{figure:group_numerical_example}
\end{figure}

\subsubsection*{Group closeness centrality}

According to Definition~\ref{definition:spec_closeness}, each node outside a group $S$ evaluates each node in $S$ by the negative of the distance between them.
The resulting group extensions correspond to the exclusive variants discussed in Example~\ref{example:group_intro}.

Under the \textsc{Sum}, \textsc{Max}, and \textsc{Min} aggregations, the group score is the reciprocal of the sum, over all nodes outside the group, of (1) the sum of distances to all group members, (2) the distance to the closest group member, and (3) the distance to the farthest group member, respectively.
These definitions correspond, up to normalization, to the mean, minimum, and maximum variants of group closeness introduced by \citet{Everett:Borgatti:1999}.
Specifically, our \textsc{Sum} variant corresponds to the mean variant, but uses the sum of distances instead of their average over the group members.
Our \textsc{Max} variant corresponds to the minimum-distance variant, since maximizing the negative distance selects the closest group member.
Conversely, our \textsc{Min} variant corresponds to the maximum-distance variant.

The group score obtained under the \textsc{Prop} aggregation lies between those obtained under the \textsc{Sum} and \textsc{Max} aggregations.
All distances from nodes outside the group to its members contribute to the score, but larger distances receive progressively smaller weights.

\subsubsection*{Group betweenness centrality}
According to Definition~\ref{definition:spec_betwennness}, each pair of distinct nodes $s,t$ outside a group $S$ evaluates a node $v\in S$ by the fraction of shortest paths between them that pass through $v$.
If no path connects $s$ and $t$, the evaluation is zero.

Under the \textsc{Sum} aggregation, each pair sums its evaluations of all group members.
Thus, for groups of fixed size, the group value is proportional to the expected number of group members encountered on a uniformly chosen shortest path between a uniformly chosen pair of distinct nodes outside $S$.

Under the \textsc{Max} aggregation, each pair retains only the highest evaluation.
This means that for a fixed pair $\{s,t\}$, only the largest fraction of shortest paths passing through a single group member is counted.
Importantly, this is not necessarily the fraction of shortest paths passing through at least one group member, as different paths may pass through different members.

In turn, under the \textsc{Min} aggregation, each pair retains the lowest evaluation.
Hence, this evaluation is positive only if every group member lies on at least one shortest path between the evaluating pair, although not necessarily on the same path.

Similarly to group closeness, the group value obtained under the \textsc{Prop} aggregation lies between those obtained under \textsc{Sum} and \textsc{Max}.
It accounts for the contributions of all group members, but assigns progressively smaller weights to lower evaluations.

We note that these definitions differ from the classical group betweenness centrality of \citet{Everett:Borgatti:1999}, defined as $F_S(G)=\sum_{\{s,t\}\subseteq V\setminus S,\ s\neq t}\nicefrac{\sigma_{st}(S)}{\sigma_{st}}$,
where $\sigma_{st}(S)$ denotes the number of shortest paths between $s$ and $t$ that pass through at least one member of $S$.
This definition cannot be recovered from the above SPEC representation of betweenness centrality.
The reason is that each evaluating pair assesses the members of $S$ separately, and therefore cannot determine whether different group members lie on the same shortest path or on different shortest paths.
For example, the pair $\{1,6\}$ cannot distinguish between the groups $\{2,3\}$ and $\{2,4\}$ in
\tikz[baseline=-0.5ex, x=0.45cm,
    every node/.style={circle,draw,minimum size=3mm,
                       inner sep=0pt,font=\scriptsize}]{
    \foreach \i in {1}
        \node (n\i) at ({\i-1},0) {\i};
    \foreach \i in {2}
        \node[fill=blue!15] (n\i) at ({\i-1},0.2) {\i};
    \foreach \i in {3}
        \node (n\i) at ({\i-1},0.2) {\i};
    \foreach \i in {4}
        \node[fill=blue!15] (n\i) at ({\i-3},-0.2) {\i};
    \foreach \i in {5}
        \node (n\i) at ({\i-3},-0.2) {\i};
    \foreach \i in {6}
        \node (n\i) at ({\i-3},0) {\i};
    \draw (n1)--(n2);
    \draw (n2)--(n3);
    \draw (n4)--(n5);
    \draw (n5)--(n6);
    \draw (n1)--(n4);
    \draw (n3)--(n6);
},
even though only the second group intersects both shortest paths between nodes $1$ and $6$.

Nevertheless, a group measure inducing the same ranking as classical group betweenness can be obtained from a SPEC representation of endpoint-inclusive betweenness centrality, in which shortest paths, rather than pairs of nodes, serve as evaluators.
Specifically, every subset of nodes is a potential evaluator, but its evaluation is non-zero only if it forms a shortest path.
Since every shortest path in an undirected, unweighted graph is an induced path, its vertex set uniquely determines the path and its endpoints.

Formally, all subsets serve as evaluators ($\mathcal{S}(W)=2^W$), the inclusive policy is used ($P=P_{inc}$), and no conversion is required ($c(r)=r$).
The evaluation $e_G(U,v)=1/\sigma_{st}$ if $U$ is the vertex set of a shortest path with endpoints $s,t$ containing $v$, and $0$ otherwise.
For individual nodes, this representation yields endpoint-inclusive betweenness centrality, which also counts shortest paths starting or ending at the evaluated node.
Under the \textsc{Max} aggregation, each shortest path contributes $1/\sigma_{st}$ if it contains at least one group member, and $0$ otherwise.
This definition differs from classical group betweenness by also counting shortest paths whose endpoints belong to the evaluated group.
However, for connected graphs, these additional contributions depend only on the group size, so both measures induce the same ranking of groups of a fixed size.

\begin{figure*}[!t]
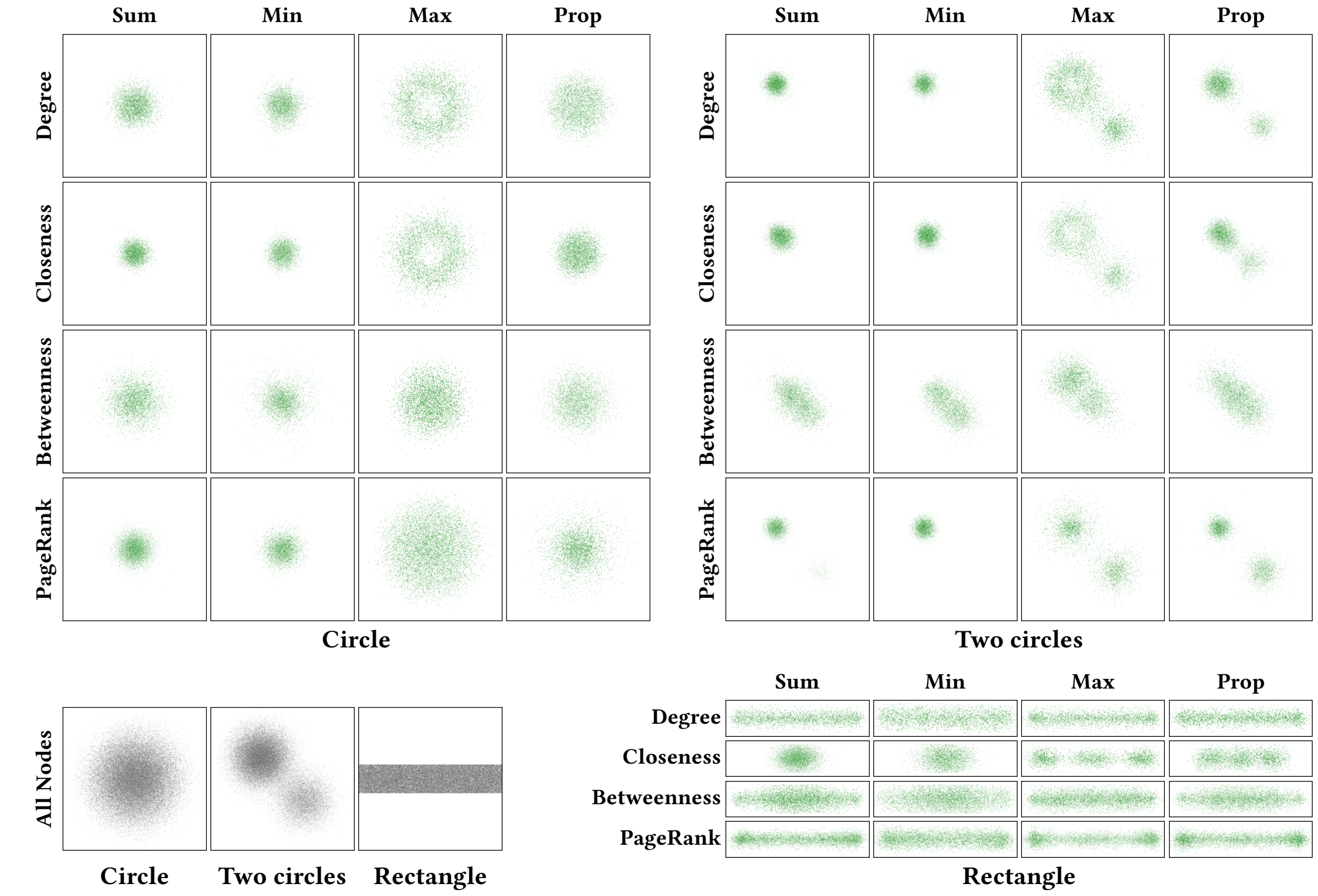

  \centering
  \PaperHeatmaps
 \caption{Spatial distributions of nodes selected in optimal groups for three random geometric graph models: Circle, Two circles, and Rectangle.
For each model, the panels show the locations of nodes selected in optimal groups across $1000$ randomly generated graphs, with darker regions indicating more frequent selection.
Rows correspond to the underlying centrality measures (degree, closeness, betweenness, and PageRank), while columns correspond to the four aggregation rules (\textsc{Sum}, \textsc{Min}, \textsc{Max}, and \textsc{Prop}).
The bottom-left panel for each model shows the underlying spatial distribution of all nodes for comparison.}

  \label{figure:experiments}
\end{figure*}

\subsubsection*{Group PageRank centrality}

According to Definition~\ref{definition:spec_pagerank}, each node $w\in V$, including nodes in the group, evaluates a node $v\in S$ by the total weighted contribution of walks from $w$ to $v$.
This evaluation is proportional to personalized PageRank, which has been used in recommender systems and to measure influence~\cite{Gupta:etal:2013,Bressan:2016}.
Thus, the evaluations can be interpreted as the relevance or influence of a node $v$ from the perspective of another node $w$.

Under the \textsc{Sum} aggregation, the group value equals the sum of the individual PageRank scores, since the inclusive policy and identity conversion are used.
Under the \textsc{Max} aggregation, each evaluator retains only the highest evaluation of a group member.
A group receives a high score when many nodes assign high relevance to at least one of its members.
An equivalent approach, up to normalization, was used by \citet{Lindgren:2016}. 

Under the \textsc{Min} aggregation, each evaluator retains only the lowest evaluation of a group member.
Thus, a node assigns a high evaluation to a group only if it considers all group members highly relevant.
Finally, under the \textsc{Prop} aggregation, all evaluations are aggregated in non-increasing order using harmonic weights.
This models situations where having multiple relevant group members is beneficial, but additional members provide diminishing returns.

We note that the Group-PageRank measure \cite{Liu:etal:2014} and several other group centrality measures can also be expressed within our framework; see Appendix~\ref{appendix:group_centralities} for details.

\section{Experiments}\label{section:experiments}
In this section, we illustrate how different group extensions of classical centrality measures behave.
We consider four centrality measures: degree, closeness, betweenness, and PageRank with damping factor $\alpha=0.85$, with their SPEC representations defined in Section~\ref{section:spec_representation}.
Each SPEC representation is extended to groups using the four aggregations: \textsc{Sum}, \textsc{Min}, \textsc{Max}, and \textsc{Prop}, resulting in sixteen group centrality measures.

To illustrate which nodes are selected in optimal groups, we generate 50-node graphs by first sampling node positions and then connecting pairs of nodes whose Euclidean distance is below a given threshold $d$.
We consider the square $[-5,5]^2$ and sample node positions according to three different spatial distributions.
Any point generated outside the square is moved to the closest point within the square.
Since closeness centrality is defined only for connected graphs, we discard any generated graph that is disconnected.

For each of the sixteen group centrality measures and each spatial distribution, we compute an optimal group of size $k$ and mark its nodes in green.
We repeat this procedure for $1000$ randomly generated graphs and superimpose the selected nodes.

The results are presented in Figure~\ref{figure:experiments}.

\subsubsection*{Circle}
In the first spatial distribution, both coordinates of each node are independently drawn from a Gaussian distribution $\mathcal{N}(0,1.5)$, producing a roughly circular cloud of nodes.
Two nodes are connected by an edge if their Euclidean distance is at most $d=1.5$.
The size of the optimal group is $k=4$.

We observe that the \textsc{Max} aggregation produces more spatially dispersed selections than the other aggregations for all four centrality measures, favoring groups in which each node in the graph has a selected node nearby.
Interestingly, for degree and closeness centralities, this happens even at the expense of selecting nodes near the center, leaving a visible gap around the origin.
For PageRank, nodes near the periphery are also selected relatively frequently.
This may be partially explained by the influence of leaf nodes, which transfer their entire PageRank contribution to their only neighbor.
In contrast, the \textsc{Sum} and \textsc{Min} aggregations predominantly select nodes near the center, with a noticeably wider spread for betweenness centrality.
The \textsc{Prop} aggregation generally produces an intermediate pattern between \textsc{Sum} and \textsc{Max}: it still favors central nodes but also selects nodes farther from the center.

\subsubsection*{Two circles}
In the second spatial distribution, three quarters of the nodes (rounded up) have both coordinates independently drawn from $\mathcal{N}(-1.5,1)$.
The remaining nodes have both coordinates drawn from $\mathcal{N}(1.5,1)$.
This produces two roughly circular clouds of nodes, one containing approximately three times as many nodes as the other.
As before, two nodes are connected by an edge if their Euclidean distance is at most $d=1.5$, and the size of groups is $k=4$.

For degree, closeness, and PageRank centralities, the \textsc{Sum} and \textsc{Min} aggregations select nodes almost exclusively from the larger cloud, where the most central nodes are concentrated.
In particular, for closeness under \textsc{Min}, selecting nodes from both clouds can significantly decrease the score, as nodes in the larger cloud must then account for the distance to a distant group member.
Under the \textsc{Prop} aggregation, some selected nodes also come from the smaller cloud, with closeness centrality showing a visible tendency to select nodes in the region connecting the two clouds.
As in the single-circle case, the \textsc{Max} aggregation favors more dispersed selections, leaving a gap near the center of the larger cloud.
For betweenness centrality, all four aggregations concentrate on nodes connecting the two clouds, with \textsc{Max} selecting more nodes from the larger cloud.

\subsubsection*{Rectangle}
In the third spatial distribution, node coordinates are drawn uniformly from the rectangle $[-5,5]\times[-2,2]$.
Two nodes are connected by an edge if their Euclidean distance is at most $d=1$, and groups of size $k=3$ are selected.

We observe that the results for closeness centrality resemble those from Example~\ref{example:intro}, with \textsc{Sum} and \textsc{Min} favoring nodes near the center, \textsc{Max} selecting three widely separated nodes, and \textsc{Prop} producing an intermediate arrangement.
For PageRank, nodes near the ends of the rectangle are selected more frequently, particularly under the \textsc{Max} and \textsc{Prop} aggregations.
This may be partially explained by the influence of leaf nodes, which can increase the PageRank of their neighbors.
For example, in a path graph, the nodes adjacent to the endpoints have the highest PageRank.
In contrast, for betweenness centrality, selected nodes are more concentrated near the center of the rectangle.
For degree centrality, the averaged distributions resemble those for closeness, but this similarity is largely due to averaging over multiple graph instances: in individual graphs, selected nodes tend to be clustered together.

\section{Conclusions}
We developed a general framework for extending centrality measures to groups.
We proposed four extensions of each measure based on different ways of aggregating evaluations: \textsc{Sum}, \textsc{Max}, \textsc{Min}, and \textsc{Proportional}.
The framework builds on a new unified representation of classical centrality measures, which we call the SPEC representation.
Our experimental analysis showed that different variants capture different notions of an optimal group.

Our work opens up several directions for future research.
It would be interesting to study the SPEC representation axiomatically and characterize existing classes of centrality measures through restrictions on its components.
Extending existing axiomatic characterizations to group centrality measures could also shed new light on their behavior.
Finally, evaluating the proposed group centrality measures in real-world applications and assessing their practical usefulness would be a valuable next step.

\section{Acknowledgments}
This work was supported by the Polish National Science Centre Grants No. 2018/31/B/ST6/03201 and 2023/50/E/ST6/00665.

\bibliography{bibliography, additional}
\bibliographystyle{ACM-Reference-Format}

\clearpage
\appendix
\section{SPEC Representations of Centrality Measures}\label{appendix:spec}

The node centrality induced by a SPEC representation $\mathcal{R}=(\mathcal{S},P,e,c)$ is defined as
\[
F^{\mathcal{R}}_v(G)=c\left(\sum_{U\in\mathcal{S}(V\setminus P(\{v\}))}e_G(U,v)\right)
\]
for every graph $G=(V,E)$ and node $v\in V$.
In the following representations, we specify the evaluation function only for the subsets selected by the selector; for all other subsets, its value is set to zero.

Degree centrality admits the following SPEC representation:
\begin{itemize}
    \item $\mathcal{S}(W) = \{\{w\} : w \in W\}$ (individual nodes serve as evaluators);
    \item $P = P_{exc}$ (a node does not evaluate itself);
    \item $e_G(\{w\}, v) = 1$ if $\{w,v\} \in E$, and $0$ otherwise (node $v$ receives $1$ from each neighbor and $0$ from every other node);
    \item $c(r) = r$ (no conversion).
\end{itemize}

\subsection{Distance-based centralities}

Closeness centrality~\cite{Bavelas:1948} for connected graphs with at least two nodes admits the following SPEC representation:
\begin{itemize}
    \item $\mathcal{S}(W) = \{\{w\} : w \in W\}$ (individual nodes serve as evaluators);
    \item $P = P_{exc}$ (a node does not evaluate itself);
    \item $e_G(\{w\}, v) = -\dist(w,v)$ (node $w$ evaluates node $v$ by the negative of the distance between them);
    \item $c(r) = -1/r$ (converts the negative sum of distances into the reciprocal of their sum).
\end{itemize}

Harmonic centrality~\cite{Rochat:2009} admits the following SPEC representation:
\begin{itemize}
    \item $\mathcal{S}(W) = \{\{w\} : w \in W\}$ (individual nodes serve as evaluators);
    \item $P = P_{exc}$ (a node does not evaluate itself);
    \item $e_G(\{w\}, v) = \nicefrac{1}{\dist(w,v)}$, with $1/\infty=0$ (node $w$ evaluates node $v$ by the reciprocal of the distance between them);
    \item $c(r) = r$ (no conversion).
\end{itemize}

Decay centrality~\cite{Jackson:2005} with $\alpha \in (0,1)$ admits the following SPEC representation:
\begin{itemize}
    \item $\mathcal{S}(W) = \{\{w\} : w \in W\}$ (individual nodes serve as evaluators);
    \item $P = P_{exc}$ (a node does not evaluate itself);
    \item $e_G(\{w\}, v) = \alpha^{\dist(w,v)}$, with $\alpha^\infty=0$ (node $w$ evaluates node $v$ by raising $\alpha$ to the power of the distance between them);
    \item $c(r) = r$ (no conversion).
\end{itemize}

$k$-step reach centrality~\cite{Borgatti:etal:2013} with $k\geq 1$ admits the following SPEC representation:
\begin{itemize}
    \item $\mathcal{S}(W) = \{\{w\} : w \in W\}$ (individual nodes serve as evaluators);
    \item $P = P_{exc}$ (a node does not evaluate itself);
    \item $e_G(\{w\}, v) = 1$ if $\dist(w,v) \le k$, and $0$ otherwise (node $w$ assigns $1$ to nodes at distance at most $k$ and $0$ to all other nodes);
    \item $c(r) = r$ (no conversion).
\end{itemize}

\subsection{Medial centralities}

Betweenness centrality~\cite{Freeman:1977} admits the following SPEC representation:
\begin{itemize}
    \item $\mathcal{S}(W) = \{\{s,t\} \subseteq W : s \neq t\}$ (pairs of distinct nodes serve as evaluators);
    \item $P = P_{exc}$ (pairs containing the evaluated node are excluded);
    \item $e_G(\{s,t\}, v) = \sigma_{st}(v)/\sigma_{st}$ if $\sigma_{st}>0$, and $0$ otherwise (a pair of nodes $s,t$ evaluates node $v$ by the proportion of shortest paths between them that pass through $v$);
    \item $c(r) = r$ (no conversion).
\end{itemize}

Stress centrality~\cite{Shimbel:1953} admits the following SPEC representation:
\begin{itemize}
    \item $\mathcal{S}(W) = \{\{s,t\} \subseteq W : s \neq t\}$ (pairs of distinct nodes serve as evaluators);
    \item $P = P_{exc}$ (pairs containing the evaluated node are excluded);
    \item $e_G(\{s,t\}, v) = \sigma_{st}(v)$ (a pair of nodes $s,t$ evaluates node $v$ by the number of shortest paths between them that pass through $v$);
    \item $c(r) = r$ (no conversion).
\end{itemize}

Flow betweenness centrality~\cite{Freeman:etal:1991} admits the following SPEC representation:
\begin{itemize}
    \item $\mathcal{S}(W) = \{\{s,t\} \subseteq W : s \neq t\}$ (pairs of distinct nodes serve as evaluators);
    \item $P = P_{exc}$ (pairs containing the evaluated node are excluded);
    \item $e_G(\{s,t\}, v) = \flow_{s,t}(G)-\flow_{s,t}(G-v)$ (a pair of nodes $s,t$ evaluates node $v$ by the reduction in the maximum flow between them caused by removing $v$);
    \item $c(r) = r$ (no conversion).
\end{itemize}
Here, $\flow_{s,t}(G)$ denotes the maximum number of pairwise edge-disjoint paths between $s$ and $t$.

Random-walk counting betweenness centrality~\cite{Bloechl:etal:2011} admits the following SPEC representation for connected graphs with at least two nodes:
\begin{itemize}
    \item $\mathcal{S}(W) = \{\{s,t\} \subseteq W : s \neq t\}$ (pairs of distinct nodes serve as evaluators);
    \item $P = P_{inc}$ (pairs $\{s,t\}$ may contain node $v$);
    \item $e_G(\{s,t\},v) = \frac{1}{|V|(|V|-1)}\sum_{k=0}^{\infty}\left(\mathbb{P}(\omega_{s,t}(k)=v)+\mathbb{P}(\omega_{t,s}(k)=v)\right)$, where $\omega_{s,t}(k)$ denotes the position at step $k$ of a random walk starting at $s$ that ends when entering $t$ (a pair of nodes $s,t$ evaluates node $v$ by the normalized expected total number of visits to $v$ during random walks in both directions);
    \item $c(r) = r$ (no conversion).
\end{itemize}

\subsection{Feedback centralities}

Beta measure~\cite{Brink:Gilles:1994} admits the following SPEC representation:
\begin{itemize}
    \item $\mathcal{S}(W) = \{\{w\} : w \in W\}$ (individual nodes serve as evaluators);
    \item $P = P_{exc}$ (a node does not evaluate itself);
    \item $e_G(\{w\}, v) = 1/\deg(w)$ if $\{w,v\} \in E$, and $0$ otherwise (node $v$ receives $1/\deg(w)$ from each neighbor and $0$ from every other node);
    \item $c(r) = r$ (no conversion).
\end{itemize}

Eigenvector centrality~\cite{Bonacich:1972} for connected graphs admits the following SPEC representation:
\begin{itemize}
    \item $\mathcal{S}(W) = \{\{w\} : w \in W\}$ (individual nodes serve as evaluators);
    \item $P = P_{inc}$ (a node also evaluates itself);
    \item $e_G(\{w\},v) = \frac{1}{|V|} \lim_{K \rightarrow \infty} \frac{1}{K} \sum_{k=0}^{K-1} \sum_{\substack{(u_0,\dots,u_k)\in\Omega(G)\\u_0=w,\ u_k=v}} (1/\lambda)^k$
    (node $w$ evaluates node $v$ by the total weighted contribution of walks from $w$ to $v$);
    \item $c(r) = r$ (no conversion).
\end{itemize}

Seeley index~\cite{Seeley:1949} for connected graphs with at least two nodes admits the following SPEC representation:
\begin{itemize}
    \item $\mathcal{S}(W) = \{\{w\} : w \in W\}$ (individual nodes serve as evaluators);
    \item $P = P_{inc}$ (a node also evaluates itself);
    \item $e_G(\{w\},v) = \frac{1}{|V|} \lim_{K \rightarrow \infty} \frac{1}{K} \sum_{k=0}^{K-1} \sum_{\substack{(u_0,\dots,u_k)\in\Omega(G)\\u_0=w,\ u_k=v}} \frac{1}{\prod_{i=0}^{k-1}\deg(u_i)}$
    (node $w$ evaluates node $v$ by its limiting average contribution of weighted walks from $w$ to $v$);
    \item $c(r) = r$ (no conversion).
\end{itemize}
For undirected connected graphs, the Seeley index is equivalent to normalized degree centrality.
We include its walk-based representation because it naturally extends to directed graphs.

Katz centrality~\cite{Katz:1953} with parameters $\alpha \in (0,1/\lambda(G))$ and $b>0$ admits the following SPEC representation:
\begin{itemize}
    \item $\mathcal{S}(W) = \{\{w\} : w \in W\}$ (individual nodes serve as evaluators);
    \item $P = P_{inc}$ (a node also evaluates itself);
    \item $e_G(\{w\},v) = b
    \sum_{k=0}^{\infty}
    \sum_{\substack{(u_0,\dots,u_k)\in\Omega(G)\\u_0=w,\ u_k=v}}
    \alpha^k$
    (node $w$ evaluates node $v$ by the total weighted contribution of walks from $w$ to $v$);
    \item $c(r) = r$ (no conversion).
\end{itemize}

PageRank~\cite{Page:etal:1999} with damping factor $\alpha\in(0,1)$, for graphs without isolated nodes, admits the following SPEC representation:
\begin{itemize}
    \item $\mathcal{S}(W) = \{\{w\} : w \in W\}$ (individual nodes serve as evaluators);
    \item $P = P_{inc}$ (a node also evaluates itself);
    \item $e_G(\{w\},v) = \frac{1-\alpha}{|V|}
    \sum_{k=0}^{\infty}
    \sum_{\substack{(u_0,\dots,u_k)\in\Omega(G)\\u_0=w,\ u_k=v}}
    \frac{\alpha^k}{\prod_{i=0}^{k-1}\deg(u_i)}$
    (node $w$ evaluates node $v$ by the total weighted contribution of walks from $w$ to $v$);
    \item $c(r) = r$ (no conversion).
\end{itemize}

\subsection{Vitality indices}

Attachment centrality~\cite{Skibski:etal:2019:attachment} admits the following SPEC representation:
\begin{itemize}
    \item $\mathcal{S}(W) = 2^W$ (all subsets of nodes serve as evaluators);
    \item $P = P_{inc}$ (evaluators containing the evaluated node are allowed, but their evaluations are set to zero);
    \item $e_G(U,v) = 2\beta(|V|,|U|)(|K(G[U])|-|K(G[U\cup\{v\}])|+1)$ if $v\notin U$, and $0$ otherwise (group $U$ evaluates node $v$ by its weighted contribution to connecting components of $G[U]$);
    \item $c(r) = r$ (no conversion).
\end{itemize}
Here, $\beta(n,k)=k!(n-k-1)!/n!$ for $0\leq k<n$, and $K(H)$ denotes the set of connected components of graph $H$.

\section{Group centralities in SPEC Representation}\label{appendix:group_centralities}
In this section we describe how existing values from the literature fit within our framework.

Group-PageRank~\cite{Liu:etal:2014}, assuming the target set $\mathcal{T}$ is $V$, admits the following SPEC representation with \textsc{Sum} aggregation:
\begin{itemize}
    \item $\mathcal{S}(W) = \{\emptyset\}\cup\{\{w\} : w \in W\}$ (the empty set and all singletons serve as evaluators);
    \item $P = P_{exc}$ (group nodes do not evaluate each other);
    \item $e_G(\{w\},v) = \frac{\alpha|V|PR_w(G)}{(1-\alpha)\deg(w)}$ if $\{w,v\}\in E$ and $0$ otherwise, $e_G(\emptyset,v)=1$ (singletons contribute their appropriately scaled PageRank worth to their neighbors and the empty set always contributes $1$, adding $S$ to the final value);
    \item $c(r) = r$ (no conversion).
\end{itemize}
Here $\alpha$ is the damping factor of Group-PageRank, and $PR_w(G)$ denotes the single-node PageRank value with damping factor $\alpha$ of node $w$.

Co-betweenness~\cite{Kolaczyk:etal:2009} admits the following SPEC representation with \textsc{Min} aggregation:
\begin{itemize}
    \item $\mathcal{S}(W) = 2^W$ (all subsets of nodes serve as evaluators);
    \item $P = P_{inc}$ (subsets containing nodes from the group also evaluate);
    \item $e_G(U,v) = \frac{1}{\sigma_{s_Ut_U}}$ if $U\in\mathcal{U}_G$, $v\in U$, $v\not=s_U$ and $v\not=t_U$ and $0$ otherwise (paths contribute one over the number of shortest paths between their ends if the node is on the path but not on its endpoints);
    \item $c(r) = r$ (no conversion).
\end{itemize}
In undirected graphs with positive edge weights and no multi-edges, shortest paths up to reversal are uniquely determined by their node sets.
This is because its endpoints are the unique farthest pair in the set, and distances from either endpoint determine the order of its nodes.
Here, $\mathcal{U}_G$ is the family of these node sets, and $s_U,t_U$ denote start and end of the path associated with $U\in\mathcal{U}_G$.

M-reach \cite{Borgatti:2006} admits the following SPEC representation with \textsc{Max} aggregation:
\begin{itemize}
    \item $\mathcal{S}(W) = \{\{w\} : w \in W\}$ (singletons serve as evaluators);
    \item $P = P_{exc}$ (group nodes do not assess each other);
    \item $e_G(\{w\},v) = 1$ if $\dist(w,v)\leq m$ and $0$ otherwise (node $v$ receives $1$ from all nodes not further than $m$);
    \item $c(r) = r$ (no conversion).
\end{itemize}
Here, $m$ is a parameter of the value.

\clearpage
\end{document}